# Automated Validation of COBOL to Java Transformation

Atul Kumar[1], Diptikalyan Saha[1], Toshikai Yasue[2], Kohichi Ono[2], Saravanan Krishnan[1], Sandeep Hans[1], Fumiko Satoh[2], Gerald Mitchell[3], Sachin Kumar[3]

[1]IBM Research - India, [2]IBM Research - Tokyo, [3]IBM Software
kumar.atul@in.ibm.com

## ABSTRACT

Recent advances in Large Language Model (LLM) based Generative AI techniques have made it feasible to translate enterprise-level code from legacy languages such as COBOL to modern languages such as Java or Python. While the results of LLM-based automatic transformation are encouraging, the resulting code cannot be trusted to correctly translate the original code. We propose a framework and a tool to help validate the equivalence of COBOL and translated Java. The results can also help repair the code if there are some issues and provide feedback to the AI model to improve. We have developed a symbolic-execution-based test generation to automatically generate unit tests for the source COBOL programs which also mocks the external resource calls. We generate equivalent JUnit test cases with equivalent mocking as COBOL and run them to check semantic equivalence between original and translated programs. Demo Video: https://youtu.be/aqF_agNP-lU

## CCS CONCEPTS

• **Software and its engineering** → **Software testing and debugging**.

## KEYWORDS

Automatic Validation, COBOL to Java, External Resource Testing

## 1 INTRODUCTION

Transforming legacy applications to some modern programming languages and platforms is an important problem for major businesses. While there exist solutions that use classical program analysis-based techniques, they do not scale at large code base and the translated code is often not easy to understand and maintain. In past few years, advances in Large Language Model (LLM) based translation tools have made it possible to generate natural looking easy to understand code. But they come with their own problems of accuracy, especially for legacy languages.

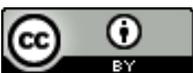

We present a tool and framework that helps in validating the equivalence between the original COBOL code and the translated Java code. The tool first generates test inputs for the COBOL code automatically. These tests cannot be run as is because of the absence of external resources (eg, databases, middleware etc) in the dev environment. Therefore, external resource calls are identified and input values for them are also generated. These tests are executed on a mainframe system with appropriate mocking for external resource calls. Output values generated by the COBOL test execution are then used to generate Java unit test assertions. External resources are equivalently mocked in Junit test cases. Automatically generated assertions compare the test output from COBOL test execution with Java test execution along multiple paths. We have implemented the tool as a Visual Studio Code extensions.

This work makes the following key contributions: 1. Automatically generate unit tests for COBOL by considering both program input variables and resource input variables for enterprise COBOL applications. 2. Automatic execution of COBOL unit tests on a mainframe system with mocking for external resources. And, 3. Automatic generation of Junit test with equivalent resource mocking. Assertions are also generated using the COBOL test outputs and variable mappings.

## 2 RELATED WORK

To the best of our knowledge, we have not come across any product, tool or research that can use automatically generated unit tests to validate the equivalence of two programs across programming languages and platforms/middleware. Some early work in validating code improvement related transformation for embedded application uses equivalence check on program representation for small changes [12]. In [13], authors present a unit testing based solution to filter out invalid translations. They leveraged automatic unit test generating tools such as Evosuite (Java) and showed promising results for Java to Python and Python to C++ transformations. They do not consider calls to external resources. There exist several tools and frameworks that can be used to solve parts of the overall problem our system tackles. COBOL-Check [1] works as a unit testing framework such as JUnit for Java and generates a new COBOL program from test case and the original program which then can be compiled and executed. This framework expects external resources to be setup which is not a case in a dev environment. Test data generation for COBOL using symbolic execution has been attempted in [11], [10], [7] etc but they do not handle enterprise COBOL constructs like CICS, IBM, DB2 etc. A survey of Symbolic Execution based techniques can be found in [9]. There are several mocking frameworks and tools for Java such as Mockito[6], EasyMock[2], WireMock[8], JMockit[5] etc. for executing Java programs. However, existing Junit generation frameworks do not generate resource mockings.

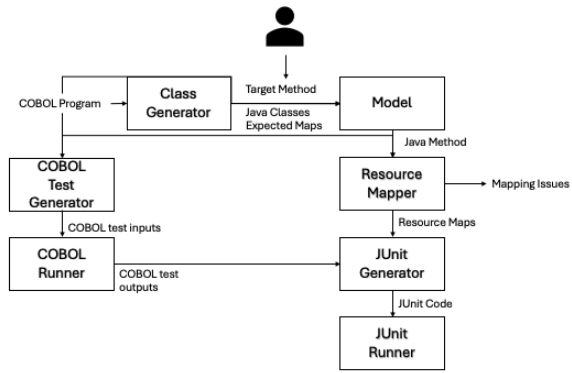

**Figure 1: Architecture**

## 3 APPROACH

### 3.1 Challenges

There exist predominantly two challenges for test-based validation of resource-centric program transformation. First, we have to ensure that the equivalence of output is checked on the maximum number of paths in the program. Therefore, we have to develop a scalable test generation algorithm to maximize coverage. Second, due to the lack of resource setup, equivalent mocking needs to be performed across both programs. This entails creating a resource map that maps each variable in COBOL resource calls (e.g. SQL SELECT) to parameters in a Java function call sequence (e.g. JDBC PrepareStatement, executeQuery, get).

We develop the dynamic validation framework to support these challenges by leveraging the branch-coverage-based symbolic execution to generate a set of appropriate test data, not only for program inputs but also for resource inputs. Subsequently, we leverage the model training data to automatically determine COBOL and Java statement mapping patterns to map resource statements and their parameters.

### 3.2 Architecture

The tool's architecture is shown in Figure 1. The input to the system is a COBOL program. The class generator creates a set of Java classes along with their member variables and method signatures by analyzing the various record structures and paragraphs present in the COBOL program. It determines how each COBOL variable will be created in Java (as a local variable, method parameter, or class variable) and creates a map (called CJMap) between COBOL and Java elements: records to classes, paragraphs to functions, and COBOL variables to the three forms of Java variables. The model uses this variable mapping and a COBOL paragraph to generate the Java function body coherent with the rest of the classes, which is then incorporated into the code with the necessary library imports. The class generator and model are beyond the scope of this paper and are not discussed any further.

The remaining components aim to validate the functional equivalence between the COBOL paragraph and the compilable Java function by generating test cases and executing them with the necessary mocking. The COBOL test generator statically analyzes the COBOL program to generate values for input variables and a list of output variables. The COBOL test runner uses these test inputs to run the program without interacting with any resources, collecting the test outputs. The resource mapper maps the COBOL external call arguments to the Java call arguments to enable equivalent mocking. The JUnit generator component produces JUnit code that initializes the Java inputs using COBOL inputs and the variable and resource maps. It also creates assertions for the Java variables, equating their values with the corresponding COBOL output variable values.

### 3.3 COBOL Test Generation

We use Symbolic execution based unit test generation techniques [9] with our proprietary algorithms for COBOL data and statement modelling and optimizations on constraint solver to support enterprise grade programs. The first step in test generation is to translate the target COBOL program to a proprietary Intermediate Representation (IR). The IR is defined as a primitive imperative language, independent of the programming languages, to represent information related to test data generation, such as control flow, data flow, and other programming constructs. Then execution paths are enumerated from the target program by considering test coverage. Selection strategy of branch direction is basically random, but using some heuristics to weight each branch edge calculated by the number of the succeeding uncovered branches. As a result, a set of enumerated execution paths might be redundant set (ie., it is not guaranteed that the number of test input data in the set is minimum). For each identified execution path, every updated variable is renamed by using SSA (Static Single Assignment). Execution condition of the execution path is calculated by symbolic execution technique. The calculated execution condition of the execution path is passed to Constraint Solver as a constraint logic formula in order to find the satisfiable solution of the constraint. As a Constraint Solver, an external SMT (Satisfiability Modulo Theories) Solver is used. The values in the satisfiable solution of the constraint logic formula are the test input data that reproduce the run of the execution path. The values for input parameters of the target program that satisfy the execution condition are calculated by using Constraint Solver. One set of test data is generated for each path.

To keep reliability of the results, self-verification is implemented to verify the generated test data. After generating the test data, we execute Symbolic Execution again with the generated test data, which is used to feed the input value for the parameters of the program and each resource. The verification is performed whether all the IR operations are performed by constant values and whether the executed path is equivalent to that for the test data.

### 3.4 Debugger-based Execution

COBOL Debugger-based program execution allows for precise manipulation of the program points. COBOL debugger allows an initial controlled setup to focus on sub-unit execution. With dynamic data inclusion at breakpoints, the COBOL debugger enables mocking through use of debug step execution data substitution and exclusion in statement step-over. This allows for application of generated data to the program within the conditions necessary for correct output values to be created and recorded correlating to the program execution pattern of the data without COBOL external calls to real systems or virtual substitution mechanisms.

## 3.5 Resource Mapping

COBOL external calls essentially refer to statements that perform resource operations (SQL statements, file processing statements) or make external calls (calls to other paragraphs and programs) outside the target paragraph. Matching COBOL external calls to Java function calls is straightforward, as they are expected to follow the CJMap. However, matching COBOL resource calls and their parameters to Java function calls and their arguments is challenging, especially in the presence of multiple identical calls, missing calls, and equivalent Java translations.

The Resource Mapper component has two steps:

- Offline Stage: It creates a map, called CJResourceMap, between COBOL and Java resource calls by automatically analyzing the COBOL-Java pairs used in model training.
- Online Stage: It first matches each COBOL resource call to potentially a set of Java sequences using CJResourceMap and provides weights representing the degree of the match. Subsequently, it solves an optimization problem that maximizes the weighted sum and performs diversified matching. For instance, if there are two identical COBOL statements and two matching Java sequences, the preferable diversified matching should not map both COBOL statements to a single Java sequence. The relative position of COBOL statements should be respected in Java, as transformations typically do not change the order. The component generates a map of COBOL variables to Java call arguments and any COBOL statements that did not match any Java Statements.

## 3.6 JUnit Generation

The tool generates JUnit test code to validate the generated Java functions corresponding to each COBOL test, utilizing the COBOL inputs and outputs. This process involves the creation of initialization code, where local and class-level variables are initialized using the COBOL-Java variable mapping and COBOL program inputs.

Based on CJResourceMap, the tool first identifies the Java external library calls corresponding to COBOL resource statements, resource input (values received by the program from the resource), and resource output variables (values going to the resource from the program) among the parameters and return variables. It then mocks the statement, to 1) gather the resource output variable values which are compared against corresponding COBOL resource output in equality-based assertions, and 2) supply the COBOL resource input values to the mapped Java resource input variables.

The mocking is made order-dependent which means if multiple instances of the same call appear in the program, then the corresponding set of values is provided from COBOL resource inputs values gathered during COBOL program execution.

Finally, it generates assertions to match the COBOL and Java program outputs.

## 4 TOOL FLOW

Validation tool is a part of IBM watsonx Code Assistant for Z (WCA4Z) [4]. WCA4Z is available as a Visual Studio Code (VS Code) extension. The validation module can be triggered from the right click context menu on a COBOL program in the VS Code after the automatic transformation Java has been performed already. The validation tool starts by parsing the COBOL program and generates an Abstract Syntax Tree (AST) using the IBM Z Language Server (available in VS Code via an extension). The AST is passed to the COBOL Unit test generation module that generates the unit tests for each paragraph of the given COBOL program. The COBOL unit tests also include inputs for the external resources. Generated unit tests are then used to create debug scripts for IBM Z Mainframe system. Debug scripts for running unit tests on an actual IBM Z Mainframe system (on cloud) are submitted as mainframe jobs. The program output and resource output are automatically captured for all tests executed. Next, the transformed Java program from the Java project in VS Code is selected and validation is invoked from the right click context menu. This triggers automatic generation of JUnit test cases. The system runs JUnit tests using the mocking libraries (provided as jar libraries) for all mainframe specific systems and functions (eg CICS, IMS etc). The tests are executed automatically using JUnit and a report is generated containing pass or fail information for each test.

## 5 EXPERIMENTAL VALIDATION

We used nine programs from different applications to run the complete validation pipeline. These applications are chosen so that they cover a range of COBOL statements and also the calls to external resources such as databases and middleware/transaction frameworks. These programs are selected from the following applications:

(1) **Genapp**[3]: an IBM demonstration mainframe application for insurance domain. This is an interactive/online application that uses CICS for transaction processing and DB2 databases as well as VSAM files for data storage.
(2) **IMS Bank**: IMS Bank is another IBM demonstration mainframe application. This application uses IBM's IMS Database.

## 5.1 Experimental Setup

For each program, we selected one COBOL paragraph with maximum resource interaction for translation. The translation is carried out by the method described in Section 4. These programs have 300-1500 line of code. For each COBOL paragraph, we manually gathered the total number of paths and branches and computed the coverage by looking at the generated test inputs. We also gathered the statistics related to generated inputs and assertions. The next subsection presents the details.

## 5.2 Results

We showcase the results of two stages (COBOL test case generation and Java JUnit generation) in table 1 with 9 representative COBOL programs with one paragraph. We have manually computed the number of branches and paths in our test program for verification. There are 4 COBOL test cases generated for 'LGACDB01' (row number 7 in table 1) with 10 program output variables and 6 resource output variables. At the Java JUnit generation part, there are 22 assertions created (3 for program output variables and 19 for resource output variables). There are 4 JUnit tests created and all of them pass the tests that validates the transformation. There are

| S. No | Program | Paragraph/Method | % Paths covered | % Branches covered | # Paths /tests | # Prog. output | # Res. output | # Assertions | # Prog. assertions | # Res. Assertions | # Tests Pass |
|---|---|---|---|---|---|---|---|---|---|---|---|
| 1 | CHANN11 | FIRST-SENTENCE | 100 | 100 | 2 | 21 | 8 | 9 | 5 | 4 | 0/2 |
| 2 | ICDBGHNP | GET-CUSTACC2 | 100 | 100 | 5 | 42 | 34 | 7 | 7 | 0 | 5/5 |
| 3 | ICDBGNP | GET-CUSTACC2 | 100 | 100 | 5 | 37 | 34 | 7 | 7 | 0 | 5/5 |
| 4 | ICGCUDAT | GET-CUSTOMER-DATA | 100 | 100 | 3 | 16 | 15 | 3 | 3 | 0 | 3/3 |
| 5 | ICLOGOUT | LOGOUT | 100 | 100 | 4 | 28 | 24 | 6 | 6 | 0 | 4/4 |
| 6 | ICSCUDAT | BEGIN | 100 | 100 | 1 | 3 | 4 | 0 | 0 | 0 | 4/4 |
| 7 | LGACDB01 | INSERT-CUSTOMER | 100 | 100 | 4 | 11 | 6 | 22 | 3 | 19 | 4/4 |
| 8 | LGAPDB01 | INSERT-POLICY | 100 | 100 | 3 | 19 | 21 | 7 | 7 | 0 | 2/2 |
| 9 | LGIPVS01 | FIRST-SENTENCE | 100 | 100 | 2 | 33 | 5 | 3 | 3 | 0 | 2/2 |

Table 1: Experimental results

2 COBOL test cases generated for 'CHANN11' (first row in table 1) with 21 program output variables and 8 resource output variables. At the Java JUnit generation part, there are 5 assertions created for the program variables and 4 assertions created for resource variables. There are 2 JUnit tests are created and both of them failed. On examining the values of the test input variables we could see that the following COBOL code block

```
PERFORM VARYING WS-CNT FROM 1 BY 1
        UNTIL WS-CNT > WS-LOOP-ITERATIONS OR
            WS-EXIT-EARLY = 'Y'
```

has been transformed incorrectly as the Java code block below:

```
while (wsCnt <= wsLoopIterations && wsExitEarly == 'N')
```

Note that the loop condition is incorrectly inverted when the condition for UNTIL in COBOL is converted to the condition for while in Java. It should be:

```
wsCnt <= wsLoopIterations && wsExitEarly != 'Y'
```

Note that the number of java test cases generated may not be equal to the COBOL test inputs. Consider 'ICSCUDAT' (row number 6 in table 1) that has only one COBOL test case while it has 4 java test cases. All the branches of the COBOL code is part of a PERFORM block and hence one cyclic path is sufficient to capture them all. Also, note that the number of assertions generated may not match the COBOL program outputs. This occurs because the COBOL test generation module might check the value of a variable, but during transformation, the variable may become a local variable in Java. Consequently, these local variable values cannot be asserted in JUnit tests, making it difficult to replicate the precise variable checks from COBOL.

The result demonstrates that our tool can successfully validate enterprise-grade COBOL transformation and able to identify potential failures in the transformation. In addition, with the scalability of the branch-coverage algorithm, we can obtain full path coverage thereby validating equivalence across all paths in the program.

## 6 CONCLUSION

We present a validation framework and tool that can be used to automatically check the semantic equivalence between a COBOL program and the automatically transformed Java program. We use automatic unit tests generation the COBOL program using symbolic execution and constraint solver techniques. This framework aims at validating enterprise applications. Therefore, we not only support enterprise COBOL programming language but also the various resource access related statements such as SQL Databases (eg IBM DB2, IMS), Transaction framework (eg, CICS, IMS) and file I/O. We use a resource input and output mechanism to solve the problem of external data access. The generated unit tests are then executed in actual IBM Z mainframe environment by automatically generating JCLs for running the tests as jobs on mainframe. We use a debugger script based approach to handle resource access related statements. Tests execution for COBOL program generates both program output and also the resource output. These along with variable mapping (which is provided by the transformation) are used to automatically generate unit tests for translated Java programs. The output values from the mainframe are used to generate assertions in JUnit programs for testing Java.

We show using some experimental results the effectiveness of COBOL test generation by looking at the branch coverage. The Java tests generation has additional metrics of number of assertion generated to validate the equivalence of COBOL and Java program by comparing the outputs.